%% file: article.tex
\documentclass[10pt,a4paper,twocolumn]{article}
\usepackage[utf8]{inputenc}
\usepackage{amsmath}
\usepackage{amsfonts}
\usepackage{amssymb}
\usepackage{makeidx}
\usepackage{graphicx}
\usepackage{lmodern}
\usepackage{kpfonts}
\usepackage[left=1.5cm,right=1.5cm,top=2cm,bottom=2cm]{geometry}
\usepackage{cite}
\usepackage{algorithmic}
\usepackage{graphicx}
\usepackage{textcomp}
\usepackage{pifont}
\usepackage{threeparttable}
\usepackage[normalem]{ulem}
\usepackage{authblk}
\usepackage[dvipsnames,svgnames]{xcolor, colortbl}
\newcommand{\etal}{{\em et al.\ }}

\definecolor{LightCyan}{rgb}{0.88,1,1}
\def\BibTeX{{\rm B\kern-.05em{\sc i\kern-.025em b}\kern-.08em
    T\kern-.1667em\lower.7ex\hbox{E}\kern-.125emX}}

\begin{document}
%
\title{Embedded Policing and Policy Enforcement based Security in the era of Digital-Physical Convergence for Next-Generation Vehicular Electronics}
\date{}
\author{Fahad Siddiqui, Matthew Hagan, Sakir Sezer}
\affil{The Centre for Secure Information Systems (CSIT),\\
Queen's University Belfast, Belfast, UK\\
(e-mail: f.siddiqui, m.hagan, s.sezer @qub.ac.uk)}
\maketitle
\section*{Abstract}
The emergence of intelligent, connected vehicles, containing complex functionality has potential to greatly benefit society by improving safety, security and efficiency of vehicular transportation. Much of this has been enabled by technological advancements in embedded system architectures, which provided opportunities for vehicle manufacturers to implement intelligent vehicle services and consolidate them within a small number of flexible and integrable domain controllers. Thus allowing for increasingly centralised operations consisting of both new and legacy functionalities. While this era of digital-physical convergence of critical and non-critical vehicle services presents advantages in terms of reducing the cost and electronic footprint of vehicular electronics, it has produced significant security and safety challenges. One approach to this research problem is to introduce fail-over mechanisms that can detect unexpected or malicious behaviours, caused by attack or malfunction, and pro-actively respond to control and minimise physical damage or safety hazards. This paper presents a novel embedded policing and policy enforcement platform architecture and the accompanied security modelling approach for next-generation in-vehicle domain controllers. To demonstrate the proposed approach, a connected vehicle case study is conducted. A realistic attack scenarios have been considered to derive security policies and enforced by the proposed security platform to provide security and safety to domain-specific features.

\input{Introduction.tex}
\input{embedded-automotive.tex}
\input{embedded-system-on-chip.tex}
\input{MPSoC.tex}
\input{threat-modelling.tex}
\input{RelatedWork.tex}
\input{SecurityModelling.tex}
\input{PolicyEngine.tex}

\input{Results.tex}
\input{connectedcar.tex}
\input{policy-enforcement}
\input{Conclusion.tex}
\bibliographystyle{IEEEtran}
\bibliography{IEEEabrv,IEEEReferences,PolicyGenerationReferences}

%




\end{document}

%% file: Introduction.tex
\section{Introduction}\label{sec:introduction}

%
%
%
%

Development of \textit{Intelligent Transportation Systems} (ITS) offers great potential to provide enhanced reliability and safety of road users, while simultaneously shortening journey times and increasing highway capacity
~\cite{Coppola2016},~\cite{Koesdwiady2016}. Developments in wireless connectivity, including \textit{IEEE 802.11p} and cellular \textit{3GPP Cellular Vehicle-to-Everything (V2X)}, have provided capability for communication between vehicles and roadside infrastructures.
Contribution of vehicle data to the cloud has enabled further capability for \textit{Data Analytics} and \textit{Machine Learning} (ML) algorithms to determine intuitive vehicle and driver information make informed decisions~\cite{Coppola2016},~\cite{Koesdwiady2016}.
\begin{figure*}[!t]
\centerline{
\includegraphics[scale=0.58]{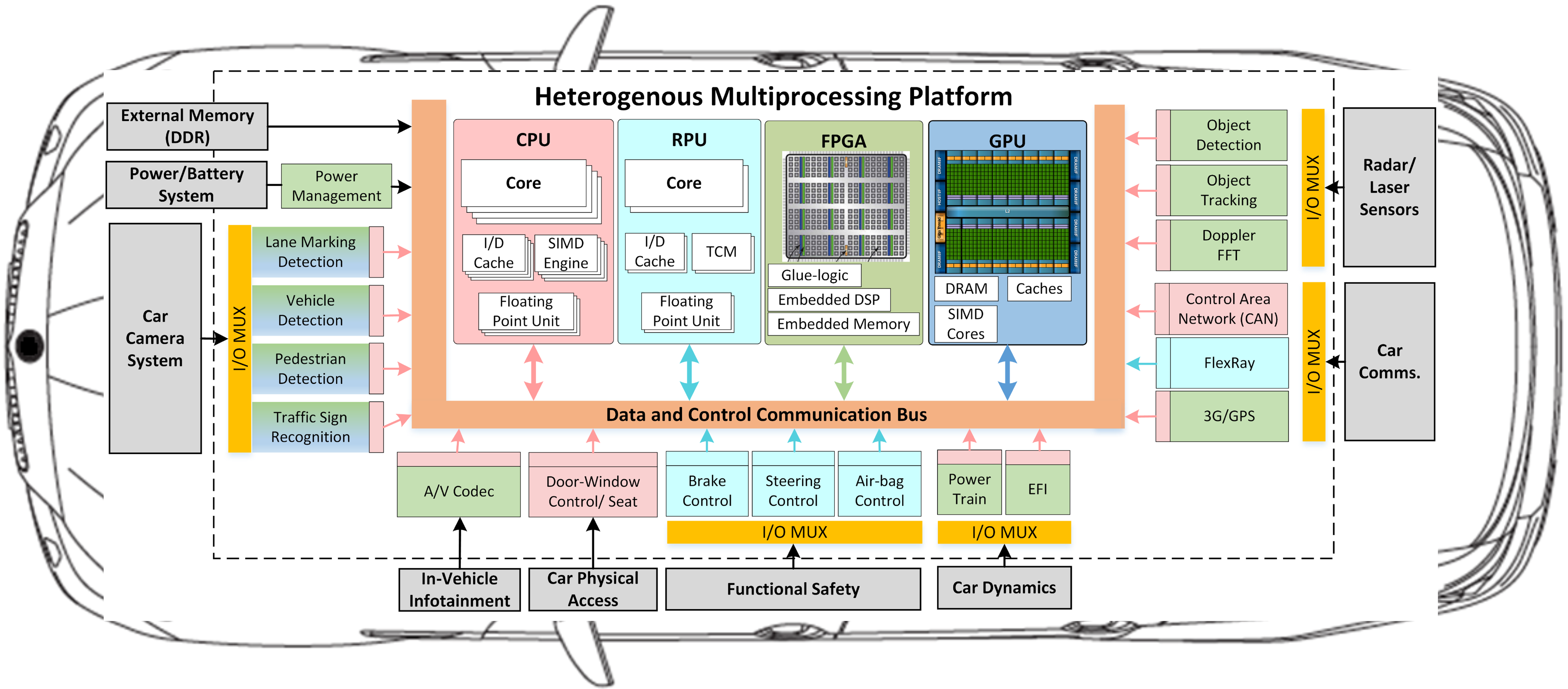}}
\caption{Block diagram view of a next-generation vehicular electronic system. Heterogeneous multiprocessing platforms offer diverse on-chip computing and power optimisation opportunities, communication protocols and architectural support to implement in-vehicle safety critical and non-critical domain controllers~\cite{XilinxAutomotive}. This approach consolidates vehicle services and subsystems into a centralised communication and control architecture.}
\label{fig:car-mpsoc}
\end{figure*}

Embedded architectures are widely used in vehicular electronics in the form of \textit{Electronic Control Units} (ECUs) that control a series of electrical and electronic systems or subsystems to ensure optimal performance, comfort and safety. They are used for both general (infotainment and navigation etc.) and safety critical (air-bag and anti-lock braking etc.) purposes. Communication among inter-connected ECUs is typically achieved using vehicle communication networks~\cite{Coppola2016},~\cite{Koscher2010}.
The normally inaccessible and closed nature of vehicular electronics has meant that until recently, securing these systems has not of significant concern to industry stakeholders, who instead prioritised safety and efficiency~\cite{Koscher2010}. Security technologies within vehicles have been confined largely to physical security, including locking, ignition, immobiliser, radio and alarm systems.
However integration of advanced embedded technologies and on-board connectivity have enabled the possibility for adversaries to carry out attacks, potentially adapted from existing attacks aimed at embedded systems,  which may enable existing crimes such as vehicle theft or unauthorised access to the vehicle's critical controls when in motion
~\cite{Koscher2010},~\cite{Woo2015}~\cite{Currie2017}.

As consumer expectations and industry standards have continued to rise, vehicle manufacturers have been faced with rising design costs due to greater system complexity. Requirements to additionally incorporate security within vehicular electronics is therefore, a challenging design and integration problem. 
Technological advancements and the emergence of \textit{Multiprocessing System-on-chip} (MPSoC) architectures have provided opportunities for vehicle manufacturers to consolidate different vehicular services into a smaller number of more flexible and integrable \textit{domain controllers}~\cite{Reinhardt2013},~\cite{ARM2019_domain}, allowing increasingly centralised operations.
However, adopting a centralised vehicular electronic system also brings a wide-range of security and safety challenges, since vehicles typically contain applications of differing levels of functional safety. Digital-physical convergence~\cite{Rajkumar2010},~\cite{Lima2016} of this mixture of critical and non-critical vehicle services onto the same device requires consideration for sufficient fail-over mechanisms to confine, control and minimise physical damage and harm to the passengers and pedestrians in event of failure or malicious event.

To assist in alleviating these issues, existing security modelling approaches used within embedded security architectures will be investigated. Our research on policing and policy enforcement architecture and security modelling approach for vehicular electronics will be introduced. Being based on the principles of access control, it provides a safety net against physical damage and harm that can be caused by a compromised system component or malfunctioning intelligent vehicle service (AI/ML), providing a means to build consumer trust in the use of intelligent services within next-generation vehicles.
This following are the novel research contributions:

\begin{enumerate}
\item A flexible, policy-based security modelling approach to alleviate the problem of defining an access-control driven security model for application use-cases with differing security requirements. The proposed approach can be integrated within existing security design practices including risk assessment and threat modelling (Section~\ref{sec:proposedpolicy}).

\item An embedded policing and policy enforcement system architecture to realise the proposed policy-based security modelling approach. The architecture monitors system bus-level communication, provides a pro-active fail-over mechanism to confine, control and minimise physical damage and harm that can be caused by malicious attack or malfunction of intelligent vehicle services (Section~\ref{sec:platform-architecture}).

\item An embedded architecture that implements a pro-active approach providing core functionalities (identify, detect, protect, respond and recover) essential for next-generation self-healing embedded systems. It detects malfunction/malicious behaviour as policy violation and pro-actively deploy fail-over mechanism to ensure safe and healthy system operation (Section~\ref{subsec:SRE}).

\item A connected vehicle use case to present the threat modelling process, that establishes critical assets and potential attacks using STRIDE to identify threat types and DREAD to quantify the potential damage of a realised attack. Policies, derived to mitigate the identified threats, are enforced by the proposed policy enforcement architecture (Section~\ref{sec:carsystem}).
\end{enumerate}

%% file: embedded-automotive.tex
\section{Background and Related Work}
\label{sec:background}
\subsection{In-vehicle Electronics}


Advancements within embedded technologies and demands to integrate intelligent features has significantly impacted vehicular electronics, increasing ECU numbers to unsustainable levels~\cite{Charette2009},~\cite{Gai2016}. ECUs control various critical and non-critical functions and are connected through vehicle communication networks and buses, including~\cite{Gai2016},~\cite{Kim2008}:

\begin{itemize}
	\item \textit{Controller Area Network} (CAN) - Multi-master based. The most widely used protocol in automobiles.
	\item \textit{Local Interconnect Network} (LIN) - Single master Polling based. Low level, low bandwidth.
	\item \textit{FlexRay} - Multi master - Redundant real-time security critical `X by wire' protocol.
	\item \textit{Automotive Ethernet} - High bandwidth, high latency protocol used for non-critical peripherals including  multimedia, Advanced Driver Assistance Systems (ADAS) cameras \& network connectivity.
\end{itemize}

To approach this problem, vehicle and original equipment manufacturers (OEM) have adopted design strategies to centralise broad groups of in-vehicle electronic features into streamlined architectures. To enable this, vehicle manufacturers have developed the concept of \textit{domain centralisation} by introducing \textit{domain controllers}~\cite{Reinhardt2013},~\cite{ARM2019_domain}. This limits the complexity that has resulted from the addition of hundreds of decision making ECUs that use different networking protocols, operating systems and software, by forming a centralised, flexible and integrable embedded architecture to control different vehicular features. This helps manufacturers simplify their designs and support growing demands for intelligent features and services~\cite{ARM2019_domain},~\cite{Sitawarin2018}.

Emerging embedded multi-core computing architectures, such as heterogeneous MPSoC, provide a mix of powerful general purpose, specialised graphics and deterministic real-time processors tightly coupled with reconfigurable logic~\cite{Siddiqui2019_J},~\cite{Amiri2017}. This presents significant design opportunities for vehicle manufacturers to explore and implement different intelligent automotive applications~\cite{Bar2016}, by partitioning and deploying multiple domain functions into a compact vehicular electronic system~\cite{XilinxAutomotive} as illustrated in Figure~\ref{fig:car-mpsoc}. Significant architectural and design benefits can be availed, such as reducing communication overheads among ECUs, providing a holistic view of the overall vehicular system, and providing cost benefits by reducing electronic footprint and wiring required within the vehicle. These architectural benefits suggest that MPSoC is a viable candidate that can allow in-vehicle \textit{domain controllers} to meet diverse application requirements of performance, safety and power.

However, adopting a centralised vehicular electronic system also brings a wide-range of security and safety challenges, since vehicles typically contain applications of differing levels of functional safety, as categorised by \textit{Automotive Safety Integrity Level} (ASIL) within the ISO 26262 standard~\cite{ISO26262}. Complex and mixed critical applications such as \textit{advanced driver assistance system} (ADAS), \textit{adaptive cruise control} (ACC), \textit{autonomous emergency braking} (AEB) and \textit{anti-lock braking system} (ABS), require real-time performance~\cite{Bar2016},~\cite{Velez2015}. Safety critical features of the vehicle, such as air-bag and braking mechanisms, have strict timing requirements, unlike non-critical components such as infotainment and window controls. The mixture of critical and non-critical vehicle services, within the same embedded system, requires consideration for sufficient fail-over mechanism~\cite{Lima2016}. This fail-over mechanism shall provide a pro-active safety net against vehicle's security and safety threats caused by compromised, malicious software/hardware components or outsider attacks. Nevertheless based on existing state-of-the-art attacks against ML models and methods, such attacks would be inevitable within vehicular electronics and would lead to life threatening consequences~\cite{Sitawarin2018},~\cite{Papernot2016}. Therefore, should a non-critical element within the vehicle fail, the reliability of any safety critical vehicular services should not be affected~\cite{Lima2016}.

To approach these vehicular electronic design and security issues, there is a need for a self healing security platform architecture for next-generation in-vehicle electronics. The architecture shall provide flexible isolation and segregation of mixed critical in-vehicle electronic services and domain controllers, a pro-active safety-net/fail-over mechanism to ensure functional and physical safety of critical services.


%% file: embedded-system-on-chip.tex
\subsection{Embedded Heterogeneous System-on-Chip}
\label{subsec:MPSoC}
\textit{Field programmable gate array} (FPGA) is a proven, reconfigurable technology that provides glue-logic to develop, optimise and implement custom hardware logic to accelerate computationally intensive applications. Emerging heterogeneous FPGA MPSoC architectures extend this appeal, providing versatile general-purpose, real-time and graphics processing computing resources (Figure~\ref{fig:car-mpsoc}), suitable to deliver the performance and power demands of next-generation intelligent data and control applications~\cite{Siddiqui2019_J},~\cite{Amiri2017},~\cite{Velez2015},~\cite{Siddiqui2014},~\cite{Shreejith2015}. Moreover, the increasing demand for intelligent and mixed criticality applications has led to integration of multiple levels of security, safety and advance power management mechanisms into new MPSoC architectures.
These architectures have extended support for the automotive industry, including ISO 26262 ASIL-C~\cite{XilinxXA2018} certification.

MPSoC architectures offer improved on-chip hardware security features to build system-level security foundation based on the principles of \textit{information assurance} and \textit{trust}~\cite{XilinxUltraScale2017}.
They facilitate realisation of a software-controlled and hardware-enforced security mechanisms that allow partitioning of system resources by dividing a system into subsystems and isolating the memory-space of these subsystem. \textit{ARM TrustZone} technology is an example, that virtually segregates hardware resources into \textit{secure} and \textit{non-secure} zones. 
These security mechanisms are extended down to the FPGA glue logic (beyond the multiprocessing cores) which ensures application and user-level security while facilitating existing hardware-software co-design practices. Such systems can run AUTOSAR compliant software stacks, bare-metal software applications or rich-operating systems such as embedded Linux. The on-chip real-time processing cores support a lock-step execution mode to implement safety critical and fault tolerant automotive applications, making the system more robust and recoverable from faulty situations, without severely degrading performance. The on-chip availability of these diverse features makes FPGA MPSoC an appealing platform for next-generation in-vehicle domain controller and vehicular electronics.

%% file: MPSoC.tex
Despite the benefits of MPSoC architectures, open literature has reported a range of software and micro-architectural vulnerabilities within system and ARM TrustZone security~\cite{Pinto2019} that have been widely used in vehicular electronics and safety-critical systems.
\begin{figure*}[!t]
\centerline{
\includegraphics[scale=0.72]{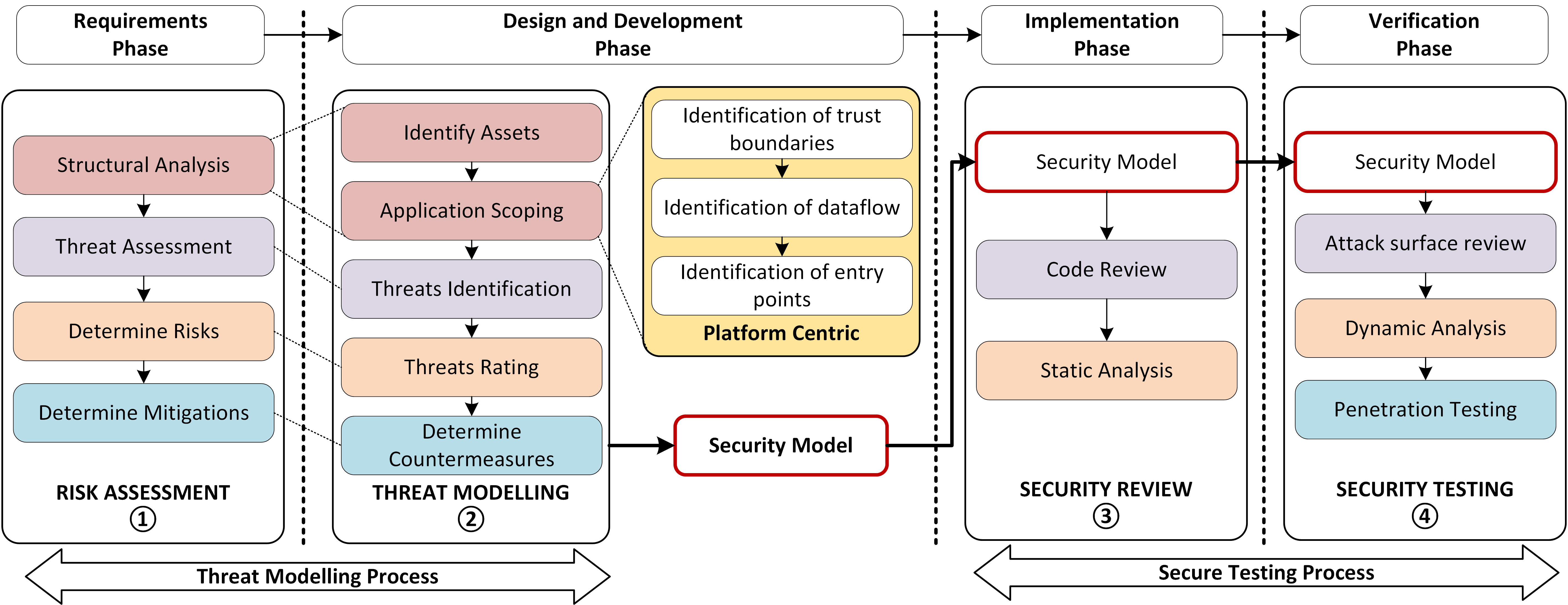}}
\caption{Step-wise illustration of secure development life-cycle covering the threat modelling and secure testing process~\cite{Howard2006}, in accordance with the different product development phases. It is important to observe that secure testing process (implementation and verification) strongly relies on the security model. Modification of the security requirements can negatively impact the implementation and verification phases.}
\label{fig:sec-model-flow}
\end{figure*}

Chen~\etal exploited a lack of proper roll-back protection during the \textit{secure boot} process used by ARM TrustZone technology~\cite{Chen2017}. They breached loading verification processes by reusing an old key. Jacob~\etal described a similar attack on the \textit{secure boot} mechanism on FPGA MPSoC, which ensures secure system start-up by building chain-of-trust~\cite{Jacob2017}. 
Mitigation of such attacks requires a well-defined methodology for distributing patches and updating keys across a wide range of devices. This is practically challenging due to differing implementations and update practices.
Ning~\etal presented a comprehensive security analysis of ARM technology, exploiting the debugging architecture and summarising security vulnerabilities and their implications~\cite{Ning2019}. 
They implemented NAILGUN attack, that exploits the debug authentication mechanism to break logical isolation, delivering privileged access to the processor's secure area.

To approach these problems, the research community has proposed various security micro-architectures that leverage execution tracing, data protection and monitoring techniques to enhance system-level security.
Arora~\etal introduced a hardware-assisted, run-time execution trace monitoring approach by embedding it deep into the processor pipeline~\cite{Arora2005}. It can detect malicious activities and safeguards the execution of programs running on embedded processors. The proposed design uses control-flow graphs to extract benign program behaviours and automatically synthesise an application-specific hardware monitor which is fixed and un-adaptable.
Maene~\etal proposed a hardware-based security architecture (Atlas) that uses memory encryption to protect instructions and cache data against system-level attack~\cite{Maene2018}, using a developed encryption unit inserted between the cache and main memory. However, this resulted in a significant performance overhead.
Coburn~\etal proposed a \textit{security-enhanced communication architecture} (SECA)~\cite{Coburn2005} that exploits bus-based system communication architectures. A wide-range of security attacks can manifest as abnormalities in system-level communication traffic. This can monitor and regulate interactions between various system components. SECA is a centralised access-control enforcement module that provides intrusion detection capabilities to the system.
On the contrary, Cotret~\etal adopted a distributed access-control security enforcement approach and proposed a \textit{hardware firewall} architecture that serves as a security bridge between the system resources and system communication bus to monitor control activities~\cite{Cotret2016}.
Though the discussed related work provides both distributed and centralised security micro-architectures to address the discussed embedded security challenges, they are limited to the detection of malicious activities. As discussed, detection alone is insufficient to meet the security and safety requirements of next-generation intelligent vehicle services and embedded architectures and must support active fail-over safety mechanisms~\cite{Siddiqui2019SoCC}.

%% file: threat-modelling.tex
\subsection{Threat and Security Modelling}
\label{subsec:application-threat-modelling}

The \textit{Secure Development Lifecycle}~\cite{Howard2006} is a process of identifying an application use case's security requirements as shown in Figure~\ref{fig:sec-model-flow}. This consists of a \textcircled{1} \textit{Risk Assessment} carried out in the requirements phase and \textcircled{2} \textit{Threat Modelling} during \textit{design and development} phase. The outcome is a \textit{security model}, a set of guidelines that aims to harnesses resilience against considered attacks. Once implemented, the \textit{secure testing process} verifies and validates the robustness of the security using a \textcircled{3} \textit{Security Review}, followed by \textcircled{4} \textit{Security Testing} to verify the implementation as shown in Figure~\ref{fig:sec-model-flow}. These steps are:

\begin{itemize}
\item \textbf{Risk assessment:}
This decomposes the use case to identify security risks within service interactions, used to formulate the system's \textit{security requirements}.

\item \textbf{Quantify Assets:}
Examined from a data flow perspective, this identifies assets requiring protection.

\item \textbf{Entry Points:}
These are system or application interfaces that may expose critical assets to an attacker.

\item \textbf{Threat Identification:}
This identifies potential vulnerabilities within the system at risk of exploitation by an attacker. STRIDE is a method that can be used to categorise identified threats within the system.

\item \textbf{Threat Severity:}
Risk-based quantification models such as DREAD can be used to evaluate risks for determined threats. This is based upon their likelihood of occurrence and severity of damage if successful.

\item \textbf{Establish countermeasures:}
Adequate countermeasures are determined based upon the previously established risk ratings.  These can be technical-based, or can be adopted through security practices.
\end{itemize}


The security modelling process faces several challenges. The implementation of the guidelines should strictly comply to the requirements as they are used for the \textit{secure application testing} phase.
Issues are likely to occur when using third-party blocks to implement dedicated tasks. Since the OEM has less influence over these components, they are constrained in mitigating potential vulnerabilities, or alternatively face delays and costs requesting modifications.

As vehicular technology has developed, from fundamental mechanics to the incorporation of electronics, software and connectivity used in embedded computing systems, so too have the methods used by adversaries, within the scope of cyber threats. Attacks against electronic vehicular systems, which have enabled unauthorised control or theft of the vehicle~~\cite{Sitawarin2018},~\cite{Valasek2015}
demonstrate a clear need to improve automotive security. Such vulnerabilities and attacks may arise from a lack of process driven security modelling, a lack of strong hardware security mechanisms or poor software implementations. As legacy automotive standards do not take into consideration the cyber domain, attempts have been made to rectify this through the introduction of the SAE J3061 standard~\cite{SAEJ30612016}, that builds upon ISO 26262's \textit{Hazard Analysis and Risk Assessment} (HARA) to define \textit{Threat Analysis and Risk Assessment} (TARA), that observes the potential for cyber security attack~\cite{ISO26262}.

While threat modelling methodologies such as STRIDE have been developed for software environments, existing work has adapted these towards physical devices and hardware. Within open literature, implementations of threat modelling methodologies have been defined for relevant use cases.
Macher~\etal~\cite{Macher2016} examined threat and risk assessment techniques integrating security and safety, presenting the application of both \textit{STRIDE} and ISO26262-HARA. Their approach, named \textit{Safety-aware hazard analysis and risk assessment} (SAHARA) was used to determine appropriate countermeasures, with an automotive battery management use case provided.
McCarthy~\etal~\cite{NHTSA2014} described a composite modelling approach for security threats to vehicles. 
They observed existing threat modelling methodologies, proposing system decomposition and creation of data flow diagrams, before identifying threats and developing use case scenarios where attacks could occur.
Dominic~\etal~\cite{Dominic2016} proposed a risk assessment framework for automated vehicles. This work assessed differences between standard and automated vehicles. Expanding on McCarthy's work
, they included threat agents to capture intentions of attackers to determine likelihood of attack. They provided an assessment, using a reference architecture and threat model.
Tan~\etal used resource isolation as a security foundation within their decentralised system-level security architecture~\cite{Tan2017}. Isolation of resources was achieved using access control-based policies.
Akatyev~\etal investigated state of the art threats against inter-connected systems~\cite{Akatyev2017}.
A holistic approach, using the STRIDE and DREAD models, was used to categorise relevant attacks and related threats.
While a detailed investigation of a real-world case study was performed, implementation of their findings into existing practises was not covered.

\begin{figure*}[!t]
\centerline{
\includegraphics[scale=0.78]{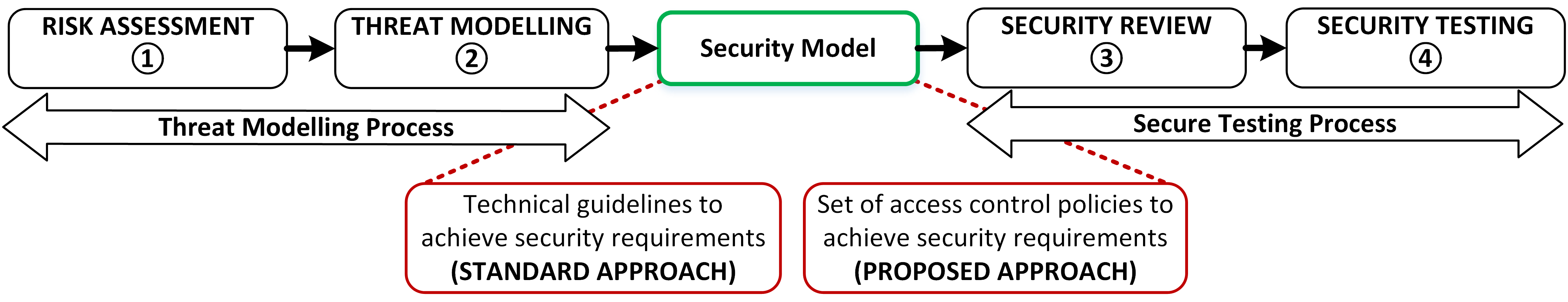}}
\caption{The security model is a bridge between threat modelling and secure testing processes. The adoption of a agile security model simplifies this process, easing modification of security requirements during the life-cycle.}
\label{fig:policy-model-flow}
\end{figure*}

The standard security modelling process faces several challenges. To be effective, implementation of the guidelines by developers should adhere to the identified requirements as they are necessary for the \textit{secure application testing} phase. The guidelines also direct developers while designing sensitive aspects the device. The use of third-party blocks to implement special or dedicated tasks is a particular risk. Since the OEM has lesser influence over these components, they are constrained in mitigating potential vulnerabilities, or alternatively face delays and costs requesting modifications to the components used.
Threats may be overlooked due to errors by the software developer or simply because they were unknown during product development. Ray~\etal~\cite{Ray2018} recognised that threat modelling is a process driven analysis that is dependent on human insights. Applying mitigation for a list of threats for each component within the device is also an onerous task~\cite{Shull2016}. In addition, should a new vulnerability be discovered, a new security model shall be devised to incorporate the change in risk profile.

To help overcome these identified issues, this work introduces a flexible and adaptable policy-based security modelling approach that can be incorporated into existing security practises. This approach compliments the published works~\cite{Ning2019},~\cite{Coburn2005},~\cite{siddiqui2018IET},~\cite{siddiqui2018},~\cite{hagan2018} and extends ~\cite{Cotret2016},~\cite{Siddiqui2019SoCC},~\cite{Akatyev2017} by presenting threat analysis of a connected vehicle case study to enforce countermeasures in the form of security policies.

%% file: RelatedWork.tex
\section{Shortcomings of existing approaches}
\label{sec:shortcomings}

The presented detailed background and related work clearly articulates that vehicle manufacturers, OEM vendors and embedded designers have faced severe security modelling and security architecture design challenges.
The following are design and architectural shortcomings identified within these existing security approaches:

\begin{itemize}
\item Heavy reliance on the principle of trust to harness and maintain security, which includes building and maintaining \textit{chain-of-trust}. The majority of embedded security micro-architectures follow \textit{Device Trust Architecture}~\cite{Global2018}. This comprises a series of nested tasks and as vulnerable as its weakest link. If broken, compromises the security of the system. In the commercial domain, \textit{Secure Boot} is a widely used secure component and found vulnerable~\cite{Chen2017}.

\item The need for compliance (in accordance with automotive cyber security standards~\cite{SAEJ30612016}), has compelled embedded security architects to design defence mechanisms that are often ad-hoc and passive in nature~\cite{Meng2018}, targeting and mitigating known class of attacks. This approach may be effective to rectify software vulnerabilities through software updates, but insufficient for micro-architecture security, as cannot be updated after release.

\item A lack of clear device security ownership, where third party components are used, insufficient adoption of security-aware design practises and an absence of baseline security requirements. In practice, design engineers do not perceive themselves accountable for security requirements and effectively embedding them into the device life-cycle, leading to security inconsistencies and vulnerabilities.

\item A lack of independent run-time monitoring and fail-over mechanism that can detect unexpected/malicious behaviours and constrain vehicle critical control to minimise physical damage and harm to the user.

%
\end{itemize}

Nevertheless, integration of vehicle electronics, that make use of AI/ML to provide intelligent services, exacerbates the risks associated with security vulnerabilities and therefore they must be handled appropriately.

%% file: SecurityModelling.tex
\section{Policy-based Embedded Security Approach}
\label{sec:proposedpolicy}

\subsection{Policy-based Security Modelling}
\label{subsec:policybasedapproach}

We propose an agile policy-based security model approach, in contrast to widely used standard guideline-based security model as shown in Figure~\ref{fig:policy-model-flow}.
This approach allows tailoring of the underlying embedded architecture to implement diverse security requirements of the application use case, without requiring redesign of the underlying architecture. This would enable automotive OEMs to customise, implement, verify, deploy and maintain different use case based critical security models, independent of silicon and third-party vendors, during the life-cycle of the system.
This can be performed, based on access-control principles, by modifying security policies at the memory and peripheral interface boundaries. These policies deploy mechanisms that physically isolate and segregate the system resources by run-time monitoring (policing) of their entry and exit points. Should a new use case be introduced, or a new vulnerability found, that changes the system's security profile, a new system security model should be generated and applied by the OEM, with updated resource specific security policies.

%% file: PolicyEngine.tex
\subsection{Policing \& policy enforcement}
\label{subsec:policing-approach}

To realise the proposed policy-based security modelling approach and alleviate the identified shortcomings, Section~\ref{sec:platform-architecture} proposes an embedded policing and policy enforcement system architecture. This architecture provides a safety net against the harm that can be caused by a compromised system component or malfunctioning intelligent vehicle service. It inspects the undergoing communications of system resources, comparing them to expected benign system behaviours. In case of unexpected behaviour caused by malfunction, malicious activity or unauthorised access by a compromised application, the system can take pro-active countermeasures as a fail-over mechanism to confine, control and minimise physical damage and harm. Expected behaviours of the system are defined as \textit{security policies}, which can be updated during the system life-cycle. 
The proposed policy-based embedded security approach is transparent and independent of existing system architectures as deployed at the system communication layer. Therefore it compliments the security extension (ARM TrustZone) of de-facto industry bus standard (ARM AMBA-AXI4). The proposed approach can be integrated across different embedded software stacks i.e. (bare-metal, embedded Linux, real-time and AUTOSAR compliant operating systems).



\section{Embedded policing and policy enforcement system architecture}
\label{sec:platform-architecture}

\begin{figure*}[h]
\centerline{
\includegraphics[scale=0.70]{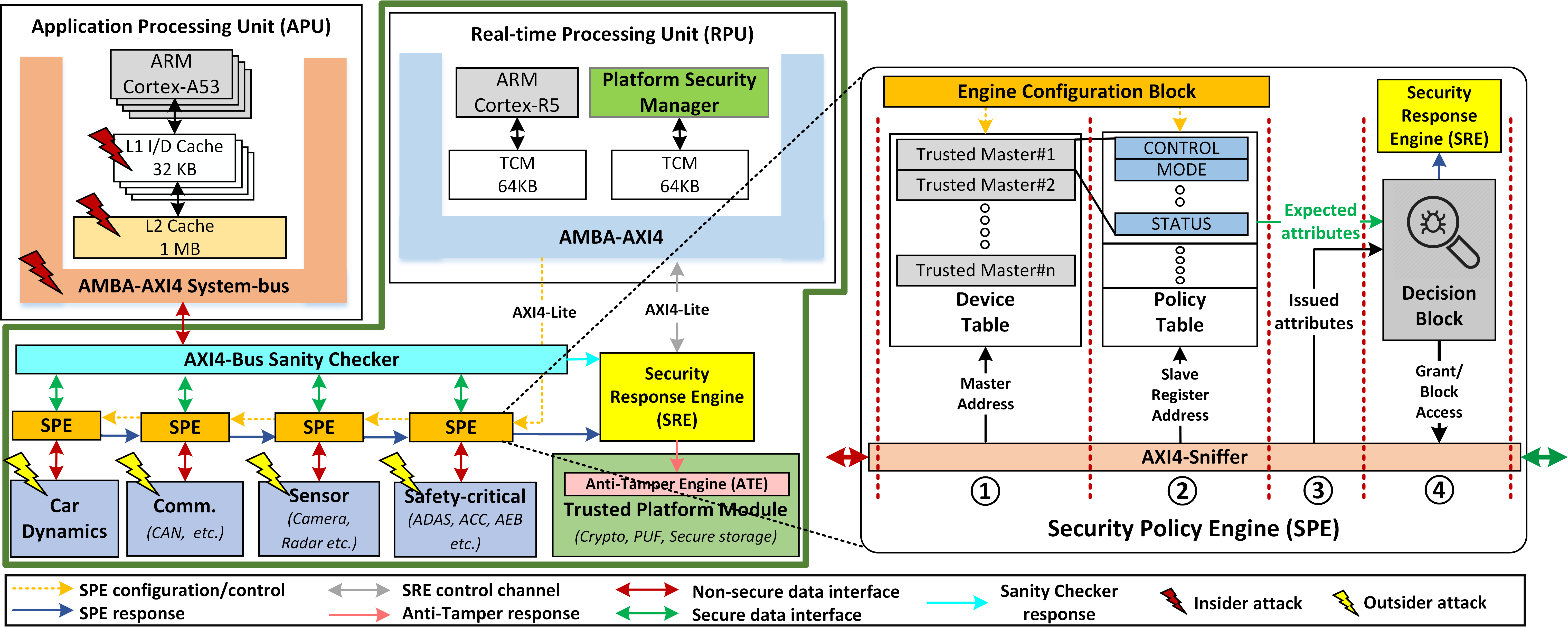}}
\caption{Block diagram of the policing and policy-based security platform architecture for next-generation in-vehicle domain controllers.}
\label{fig:platform}
\end{figure*}

To realise the proposed policing and policy enforcement approach for next-generation vehicular electronics, Figure~\ref{fig:platform} presents a system architecture that uses Xilinx Zynq UltraScale+ MPSoC platform architecture. This platform provides diverse architectural features such as re-configurable glue logic to implement and accelerate domain-specific vehicle services; on-chip hardware security features such as cryptographic primitives, \textit{Physically Unclonable Function} (PUF) and ARM TrustZone security extension to establish strong security foundation and root-of-trust.
The platform features versatile processing capabilities. It has a quad-core ARM Cortex-A53 \textit{Application Processing Unit} (APU) and a dual-core ARM Cortex-R5 \textit{Real-time Processing Unit} (RPU) as shown in Figure~\ref{fig:platform}, suitable for general purpose and deterministic time computing respectively~\cite{XilinxUltraScale2017}.

For the proposed system architecture, it is assumed that the \textit{untrusted software stack} executes on the APU and can access connected resources. This can involve execution of low risk ASIL-A or ASIL-B classified services as defined by ISO 26262. On the other hand, tasks related to embedded security and policy enforcement such as secure configuration, provisioning and management are handled by the \textit{Platform Security Manager} (i.e. one of the available \textit{Real-time Processing Units} RPUs). The relevant software stack that runs on this RPU is considered a \textit{trusted application}. In addition, the second RPU can be used to execute higher risk ASIL-C classified services, that involve time or safety critical functions, including ABS or \textit{air-bag system} control.

The APU memory is composed of two levels of cache (L1 and L2), while the RPU has a \textit{tightly-coupled memory} (TCM). Both memories are physically isolated from each other as illustrated in Figure~\ref{fig:platform}, which ensures by design that any \textit{untrusted application}  has no access to the \textit{trusted application}. This physical isolation and segregation assures that untrusted software services cannot read and modify the \textit{trusted secure application} running on the \textit{platform security manager}.
This system security boundary of the proposed architecture, highlighted in Figure~\ref{fig:platform}, provides strong physical isolation and overcomes a major shortcoming of virtualised security architectures.

\textit{Embedded policing} is an integral part of the proposed security approach, that serves as a monitor that gathers resource specific in-vehicle information. The proposed architecture leverages the bus-based communication architecture (ARM AMBA-AXI4), as a wide-range of attacks and malfunction can manifest as abnormalities within system-level communication traffic~\cite{Coburn2005}. Abnormalities and malfunctions of intelligent vehicle services are propagated down to resources as AXI4 transactions. \textit{Policy enforcement} can therefore be used to confirm whether issued transactions correspond to healthy behaviour, defined in the security policy.

According to ARM AMBA-AXI4 protocol specification~\cite{XilinxUltraScale2017}, five separate channels; \textit{address write, write data, write response, address read} and \textit{read data} are used to establish a communication link between a master and slave system components. The \textit{x}VALID and \textit{x}READY signals are used to establish master-slave handshake mechanism, while data exchange occurs when both VALID and READY signals are high. In addition, the AXI4 protocol specification supports additional control and status signals such as 3-bit AWPROT and ARPROT signal of \textit{address write} and \textit{address read} channels to incorporate system-level security~\cite{XilinxUltraScale2017}. The second bit of which is the \textit{Non-Secure} (NS)-bit that defines the security attribute of the transaction.
During system operation, these signals propagate down through the system interconnect to the FPGA glue logic to harness and maintain security which is the working concept of ARM TrustZone Security Technology. The value of the NS-bit is used to virtually segregates system resources into \textit{secure} (0) and \textit{non-secure} (1) worlds used by ARM TrustZone technology. The 2-bits RRESP and BRESP response signals of \textit{write response} and \textit{read data} channel are used to acknowledge write and read data transfers back to the host processor. The 2-bits value of these response signals (OKAY, SLVERR, DECERR) is used to inform the status back to the master, whether the data transfer transaction is successfully completed or resulted in error. However, if the NS-Bit, RRESP or BRESP signals are exposed to the compromised software or malicious hardware attack, as demonstrated in Figure~\ref{fig:ns-attacks} and Figure~\ref{fig:response-attacks}. They can be exploited to launch a man-in-the-middle, denial-of-service or slave impersonation attacks~\cite{siddiqui2018}. To realise proposed embedded policing and policy enforcement approach, the following hardware security components are essential as shown in Figure~\ref{fig:platform}:

\begin{enumerate}
	\item Security Policy Engine (SPE)
	\item Bus Sanity Checker (SCK)
	\item Anti-Tamper Engine (ATE)
	\item Security Response Engine (SRE)
\end{enumerate}

The AXI4-Lite interface allows configuration and management of each hardware security components via a \textit{trusted application}. This enables third-party vendors and security architects to explicitly provision, implement and maintain a security model during the system life-cycle~\cite{Jacob2017Provision}. This provisioning can involve actions such as explicitly disabling features and irrespective of software services running on the system. The SPE and SCK are implemented on FPGA glue logic and integrated as a hardware component, while SRE and ATE utilise hardware security features of MPSoC.


\subsection{Security Policy Engine (SPE)}
\label{subsec:SPE}
The SPE is designed to check the issued AXI4-bus transactions, compare them against a list of approved masters and their defined policies, and make a decision whether to grant or limit access to an attached slave. The Figure~\ref{fig:platform} shows a distributed deployment of SPEs that actively police the communication between the system-bus and slave peripherals including memory, sensors or actuators and enforce resource specific security measures. The SPE is a four-stage pipeline architecture as illustrated in Figure~\ref{fig:platform}, comprising of the following five hardware blocks:

\begin{enumerate}
    \item Engine Configuration Block
    \item AXI4-Sniffer
    \item Device Table
    \item Policy Table
    \item Decision Block
\end{enumerate}
\subsubsection{Engine configuration block} 
This enables secure configuration of the SPE's security parameters by the \textit{Platform Security Manager}, using the AXI4-Lite interface as shown in Figure~\ref{fig:platform}. This allows configuring and updating security policies to define resource-specific access-control rights and safe behaviour.

\begin{figure}[b]
\centerline{
\includegraphics[scale=0.62]{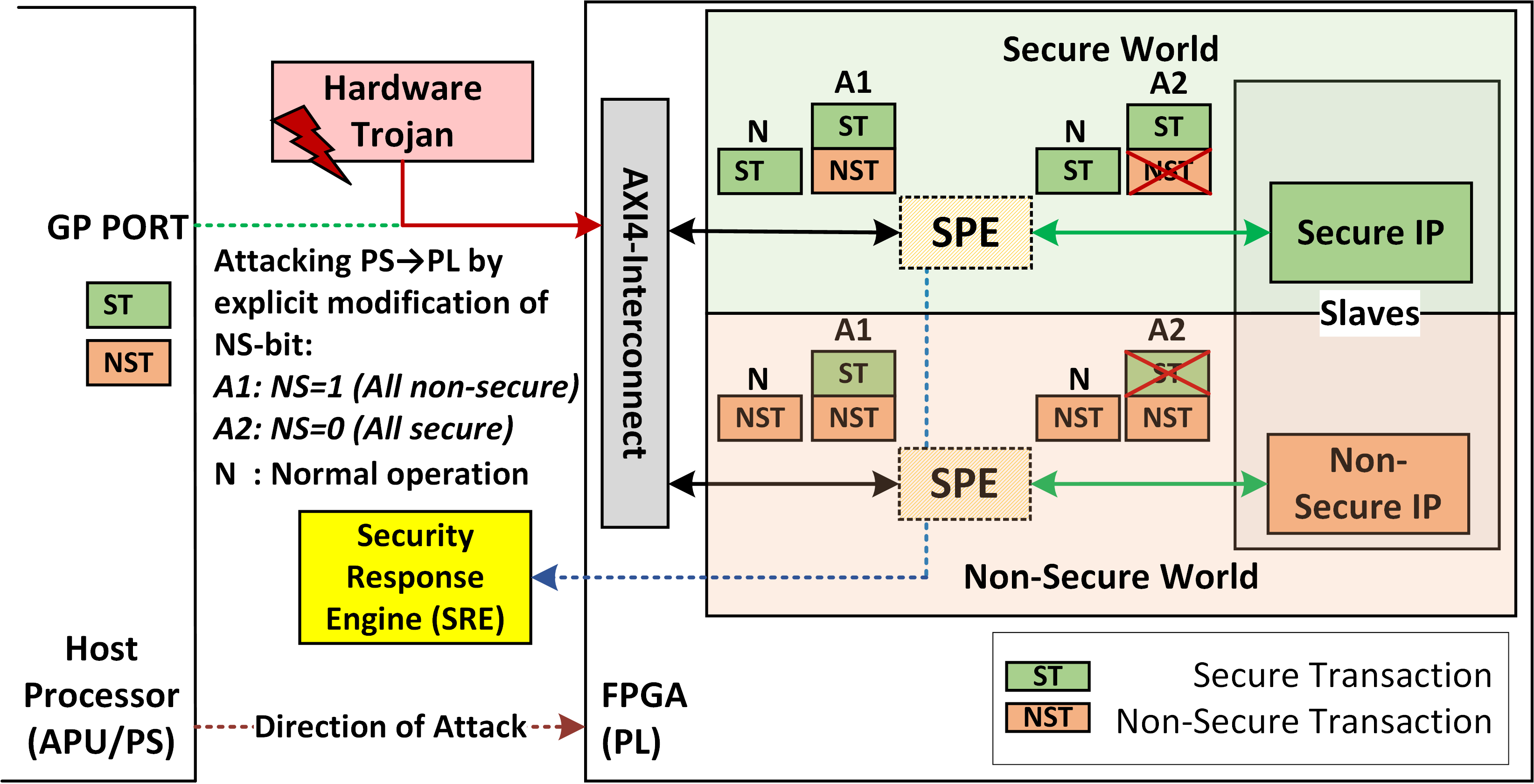}}
\caption{Logical isolation and segregation of slaves into secure and non-secure zones based on the AXI4 security attribute (NS-bit). This diagram shows the insider attack (from PS to PL) scenarios~\cite{siddiqui2018}. Moreover, how SPE can be used to detect such attacks and mitigate by the SRE.}
\label{fig:ns-attacks}
\end{figure}

\subsubsection{AXI4-Sniffer} 
The role of \textit{AXI4-Sniffer} is to ensure physical isolation between master and salve until the decision block either grants or blocks access. For this purpose, it also maintains the compulsory handshake signals to ensure correct AXI4-bus operations. It samples~\textit{master} and \textit{slave} addresses using 
Once sampled, the address is passed to the \textit{device table}, shown in Figure~\ref{fig:platform}. During this process, it deassert the \textit{x}VALID signals to the slave and \textit{x}READY signals to the master to maintain physical isolation between the master and slave.

\subsubsection{Device table} 
This contains a list of \textit{trusted masters} with access permission to slaves. Each \textit{device table} entry holds a base-address of assigned policies in \textit{policy table,} as shown in Figure.~\ref{fig:platform}. 
 
\subsubsection{Policy table} 
The \textit{policy table} contains fine-grained (register-specific) security policies for each \textit{trusted master} as shown in Figure~\ref{fig:platform}. These policies can be simple as read and write permissions to the memory and peripheral control registers, or security tailored AXI4 attributes (AxPROT).
From provisioning perspective, this allows secure and dynamic provisioning capabilities for next-generation vehicular electronics by defining policies, independent of silicon manufacturers and third-party vendors.

\subsubsection{Decision block} 
\begin{figure}[t]
\centerline{
\includegraphics[scale=0.78]{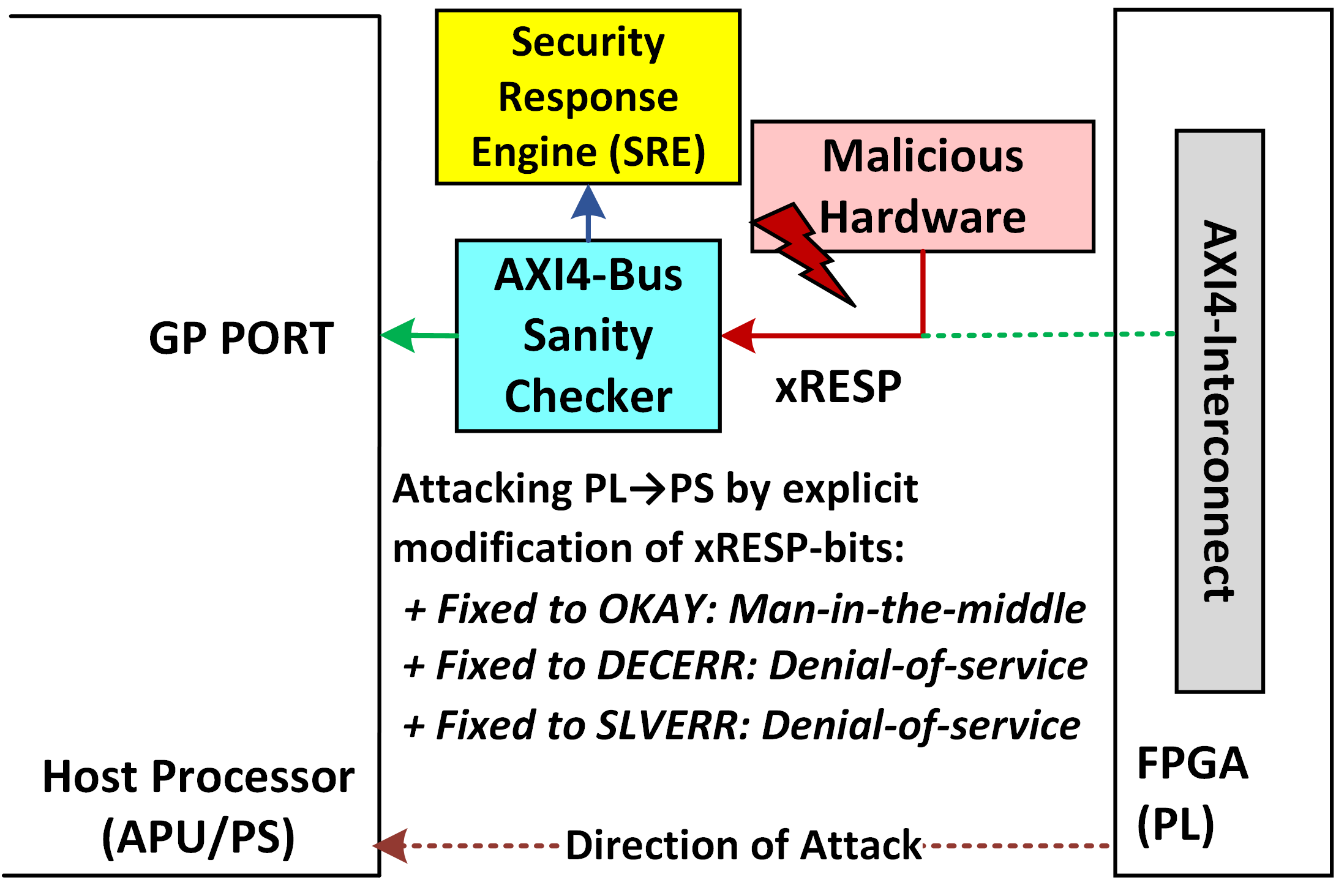}}
\caption{This block diagram illustrates an insider attack scenario (PL to PS) exploiting AXI4 write response
and read data channel signals (xRESP)~\cite{siddiqui2018}. How such attack can be detected by SCK and curtailed by SRE.}
\label{fig:response-attacks}
\end{figure}

The \textit{decision block} detects compromised service, malfunction or malicious activity, such as modification of security attributes, as shown in Figure~\ref{fig:ns-attacks} and~\ref{fig:response-attacks}. It references expected security attributes from the \textit{Policy table}, compares against issued security attributes and makes a decision as shown in Figure~\ref{fig:platform}. The result is passed to the \textit{AXI4-Sniffer} which either grants or blocks access to the slave and reporting slave's health status to SRE to initiate pro-active response.

\subsection{Bus sanity checker (SCK)}
\label{subsec:sanity}
The SCK is deployed at the interface of AMBA-AXI4 system-bus and the slaves as shown in Figure~\ref{fig:platform}, to ensure the integrity of bus response channel signals issued by slaves. These response channel signals are vulnerable and can be exploited through a compromised slave or malicious service as illustrated in Figure~\ref{fig:response-attacks}. These include main-in-the-middle and denial-of-service attacks. The SCK is designed to detect and mitigate such attacks~\cite{siddiqui2018}. SCK is composed of a three state \textit{finite-state machine} (FSM) and a programmable 32-bit timer, which is accessible only by the \textit{Platform Security Manager}. It has CONTROL and STATUS registers for enabling or disabling the peripheral, alongside reading the real-time system status reported by the SRE.

In case of (\texttt{SLVERR} or \texttt{DECERR}) attack, FSM moves from NORMAL to DETECTION state, starting a free running timer that decrements from a set value at each clock cycle. If de-assertion of \texttt{SLVERR} or \texttt{DECERR} is detected before time-out, FSM returns back to NORMAL state, otherwise it moves to ATTACK state. The ATTACK state informs the SRE to take programmed countermeasures. The \texttt{OKAY} attack is detected by comparing the status of SRE and SCK.

\subsection{Anti-tamper Engine (ATE)}
\label{subsec:ATE}
The ATE enforces system level anti-tamper countermeasures initiated by the SRE. It also provides passive physical security protection against hardware level attacks, such as side-channel analysis, by monitoring the physical parameters such as voltage and temperature. If the system hardware violates the set conditions, a system irregularity will be triggered, in the form of a system-level interrupt. The following are anti-tamper features:

\begin{itemize}
	\item Maintaining uninterrupted internal clock sources.
	\item Monitoring on single-event-upset to ensure fault tolerant execution during operational life-cycle.
	\item Voltage/temperature monitoring to ensure system operation and resist against active hardware attacks.
\end{itemize}

\subsection{Security Response Engine (SRE)}
\label{subsec:SRE}
\begin{figure}[b]
\centerline{
\includegraphics[scale=0.65]{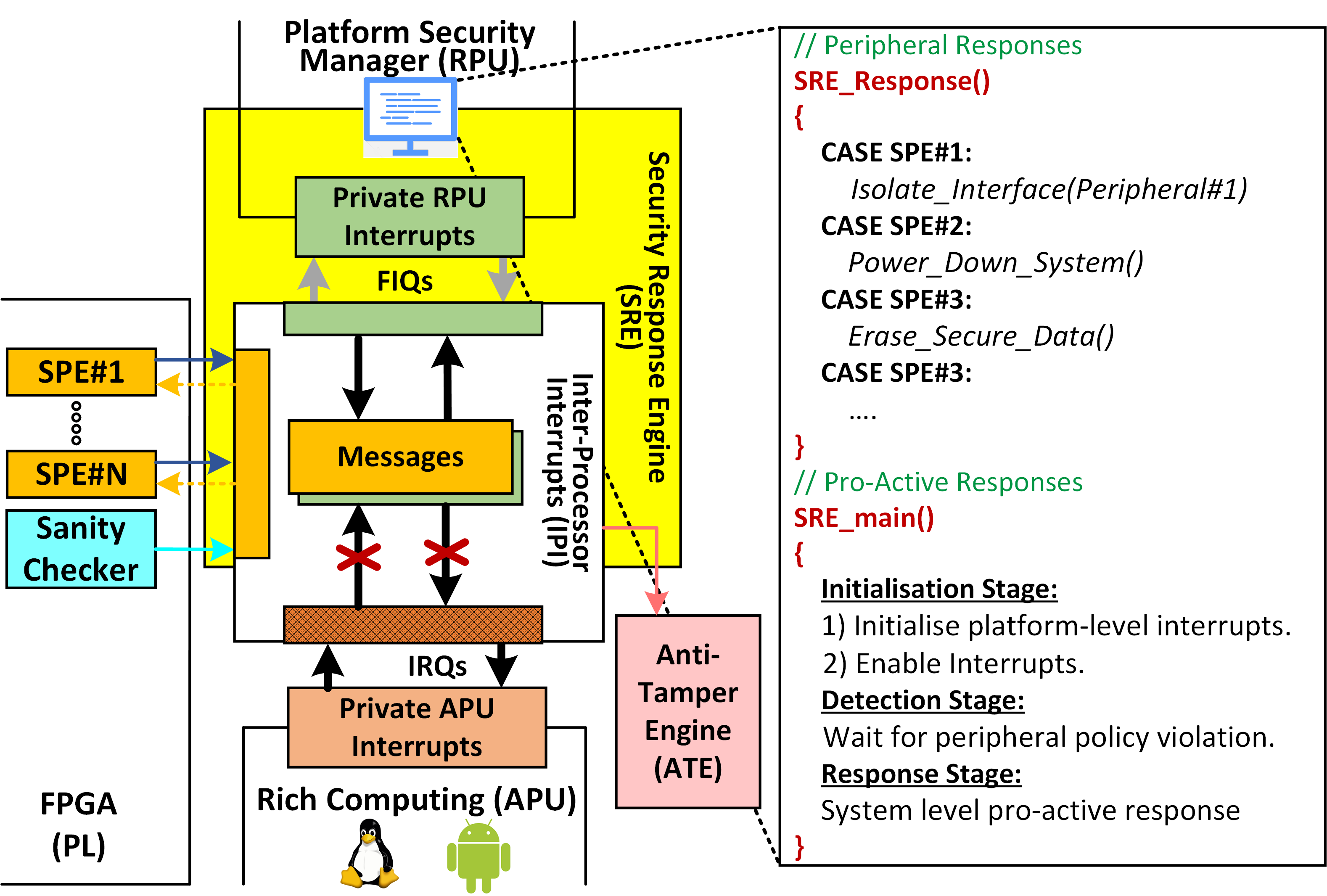}}
\caption{Block diagram representation of interrupt-based system architecture and pseudo code of Security Response Engine (SRE).}
\label{fig:sre-block}
\end{figure}

The SRE manages policy violations reported by the SPE and SCK as illustrated in Figure~\ref{fig:platform}, Figure~\ref{fig:ns-attacks} and Figure~\ref{fig:response-attacks}. It initiates and enforces programmed pro-active countermeasures against detected malicious activities. The Zynq UltraScale+ MPSoC interrupt architecture has been used, allowing communication of status, events, requests, and errors within the heterogeneous processing system as shown in Figure~\ref{fig:sre-block}. The interrupt architecture supports \textit{private processor interrupts}, for the APU and RPU, and \textit{shared inter-processor interrupts} (IPI) to exchange status, events and errors. The IPI structure allows exchange of interrupt-driven short messages using associated IPI channels with the APU, RPU and PL. These IPI channels are used for sending messages and receiving responses across the system, without the complication of autonomous read-write transactions and polling inefficiencies. In the SRE architecture, the APU interrupt channel has been disabled explicitly, as shown in Figure~\ref{fig:sre-block}, to physically prevent APU access at the architecture's security boundary.

\begin{table*}[t]
\centering
\caption{Synthesis results of proposed hardware components to realise policing approach on Zynq-7000 and Zynq UltraScale+ MPSoC.}
\label{tab:SPE-results}
\begin{tabular}{|l|l||c|cccc|c|}
\hline
\textbf{Hardware} & \textbf{MPSoC} 	& \textbf{Speed} & \multicolumn{4}{c|}{\textbf{Area}} & \textbf{Frequency} \\
\cline{5-8}
\textbf{Component} &	& \textbf{Grade} & \textbf{LUT} & \textbf{LUTRAM} & \textbf{FF} & \textbf{BRAM} & \textbf{(MHz)} \\
\hline
\hline
\textbf{SPE} & \textbf{Zynq-7000} (28nm)	& -3  & 156   & 10 & 464 & 1 & 250 \\
\cline{2-8}
 & \textbf{Zynq UltraScale+ EG} (16nm) & -3 & 185   & 10 & 464 & 1 & 300\\
\hline
\hline
\textbf{SCK} & \textbf{Zynq-7000} (28nm) & -3  & 48   & 0 & 39  & 0 & 250 \\
\cline{2-8}
 & \textbf{Zynq UltraScale+ EG} (16nm) & -3  & 56  & 0 & 39 & 0 & 333\\
 \hline
\end{tabular}
\end{table*}

The SRE architecture comprises of hardware and software components, as shown in Figure~\ref{fig:sre-block}. The hardware component uses the described interrupt architecture to receive the SPE and SCK status, in the form of policy violations at the RPU. The software component is a \textit{trusted application} that runs on the \textit{Platform Security Manager}, allowing enforcement of programmed pro-active countermeasures against detected malicious activities. For low-latency responses, the SRE uses \textit{Fast Interrupt Requests} (FIQs), which are of higher priority to normal \textit{Interrupt Requests} (IRQs). The \textit{trusted application} comprises two blocks of code as shown in Figure~\ref{fig:sre-block}. The code block \texttt{SRE\_main()} has initialisation, detection and response software functions that enable necessary interrupts, suspend normal RPU execution and execute the relevant interrupt service routines \texttt{SRE\_Response()} to enforce the defined pro-active fail-over responses. \textit{Peripheral-level responses} that isolate or deactivate specific peripheral interfaces are realised by the SPE and SCK. The ATE is used to realise \textit{system-level responses}, which can include:

\begin{itemize}
	\item Deletion of stored secret keys.
	\item Permanently disable cryptographic functions/services.
	\item Disable compromised interfaces.
	\item Initiate secure system lock-down.
	\item Initiation system reset.
\end{itemize}

From a system-level perspective, the SRE provides a holistic view of the system's resources and integrity status, detecting anomalous behaviours which would not be possible using an atomistic view. This can be used to robustly deploy system-level security measures, essential for next-generation vehicular electronics and domain controllers.

%% file: Results.tex
\section{Implementation Results}
\label{sec:validation}
The proposed architecture has been coded in Verilog HDL and simulated, synthesised and implemented using Xilinx Vivado v2018.4 on Zynq-7000 and Zynq UltraScale+ MPSoC platforms. Table~\ref{tab:SPE-results} reports the synthesis results of the developed SPE and SCK hardware security components to realise the proposed embedded  policing  and  policy  enforcement  approach. The SCK and an instance of SPE consumes less than 1\% of FPGA programmable logic resources. The device table has been implemented using Distributed RAM, utilises 10 LUTRAM, while the policy table use a BRAM, allowing storage up to 1024 policies. The timing results show that the SPE can operate at up to 250 MHz and 300 MHz on Artix-7 and Kintex UltraScale+ chips respectively. The proposed approach can be integrated within any AMBA-AXI4 compliant ASIC or SoC based embedded architecture. It can be integrated into existing and next-generation vehicular electronics and domain controllers to incorporate fail-over safety and enhance security.


%% file: connectedcar.tex
\section{Case Study: A connected Vehicle}
\label{sec:carsystem}

A \textit{connected vehicle} scenario has been undertaken as a threat and security modelling case study. This encompasses both the creation of guideline and the proposed policy-based security approaches.
Smart and connected vehicles feature an array of complex intelligent services and interconnected systems having diverse safety-critical requirements~\cite{Uhlemann2015}. 
Examples include network connectivity can be used for navigation and entertainment by the user, telemetry and firmware update by the manufacturer, or to notify emergency services in event of an accident.

Modern vehicles have been found vulnerable to attacks, with potentially serious consequences. For example, disabling of ECUs during operation has been shown possible through CAN-bus tampering. By using a malicious node to induce error messages, the selected controllers can shut off once an error threshold is reached, as per CAN protocol specification
~\cite{Palanca2017}. Furthermore, outsider attacks, launched by introducing malicious nodes to the CAN network, may further tamper with CAN signals to induce errors, causing eventual shut-down of devices~\cite{Lyu2016}.

\begin{figure}[h]
\centerline{
\includegraphics[scale=0.55]{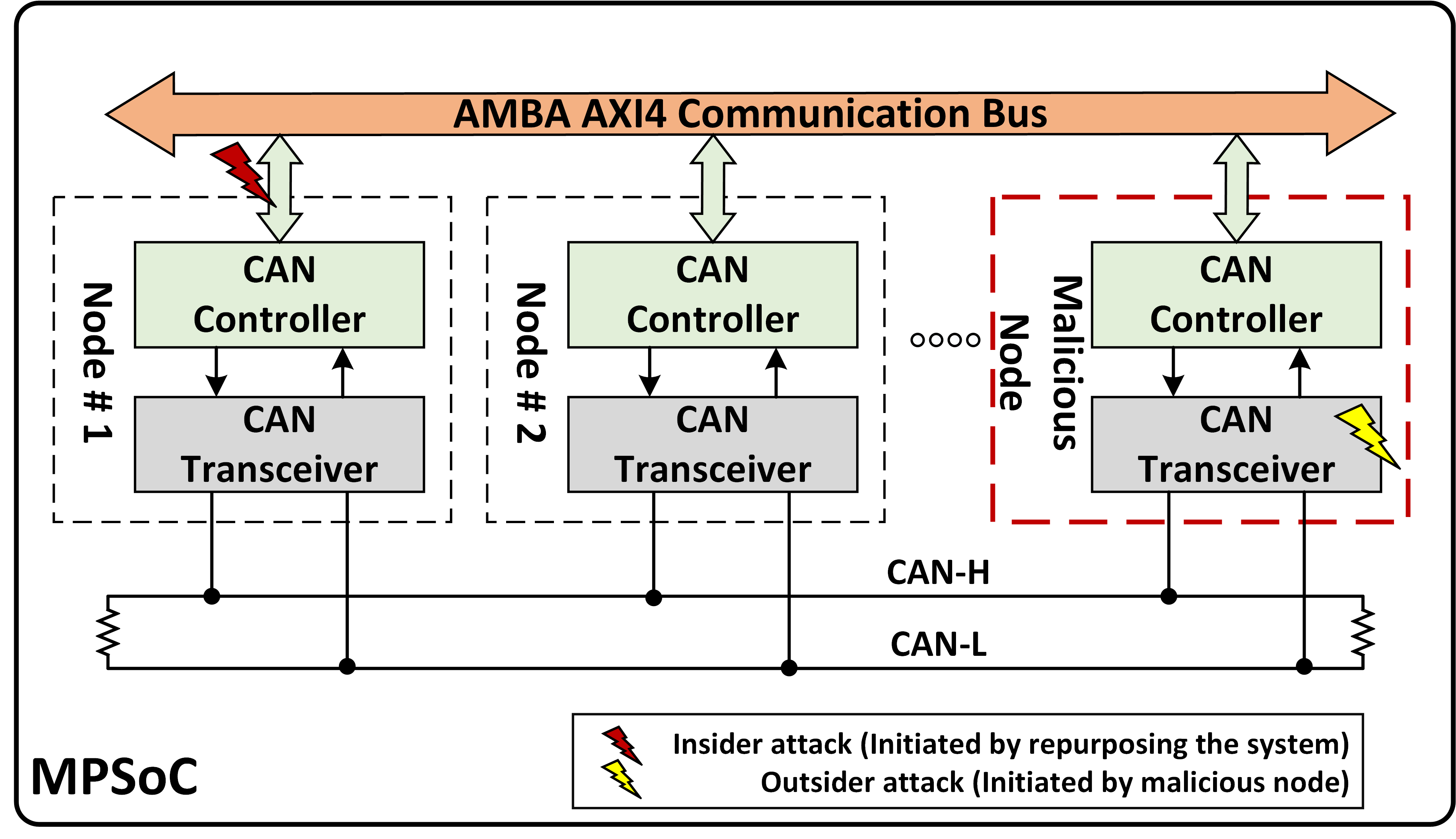}}
\caption{Block diagram of multiple CAN nodes connected via CAN bus. Each node has a dedicated CAN transceiver and controller, connected to a shared AMBA-AXI4 bus and susceptible to attacks.}
\label{fig:can-controller}
\end{figure}

For the purpose of covering the wider scope of this case study, we have considered three operation modes under which the vehicle may operate:

\begin{enumerate}
	\item \textbf{Normal:} Standard drive or park modes.
	\item \textbf{Diagnostic:} For authorised personnel.
	\item \textbf{Fail-safe:} Reserved for safety-critical situations.
\end{enumerate}

\begin{table*}[!t]
\centering
\caption{Domain-specific threat modelling of a connected vehicle. The three vehicle modes and number of relevant threats to each domain are considered. The domain critical assets are identified to derive policy-based security model of different vehicle services.}
\label{tab:app-threat-model}
\scriptsize
\begin{tabular}{|p{1.6cm}||>{\columncolor{WhiteSmoke}}c|>{\columncolor{PapayaWhip}}c|>{\columncolor{Melon}}c||p{2.9cm}|p{5.9cm}||>{\columncolor{BlanchedAlmond}}p{0.8cm}|>{\columncolor{MistyRose}}c||>{\columncolor{LightGreen}}c|}
\hline 
\textbf{Domain} & \multicolumn{3}{c||}{\textbf{Vehicle Mode}} &  &  &  & \textbf{DREAD} &  \\ 
\cline{2-4} 
\textbf{Assets} & \ding{118} & \ding{74} & \ding{89} & \multicolumn{1}{c|} {\textbf{Entry Points}} & \multicolumn{1}{c||}{\textbf{Potential Threats}} & \textbf{STRIDE} & (Avg.) & \textbf{Policy} \\ 
\hline
\hline
 & $\bullet$ &  &  & Door locks & Spoofed data over CANbus causing & STD & 8,5,4,6,4 (5.4) & R \\ 
\cline{2-5} \cline{7-9}
EV-ECU & $\bullet$ &  &  & Sensors & disablement of ECU & STD & 8,5,4,6,4 (5.4) & R \\ 
\cline{2-9} 
(accel, brake, &  & $\bullet$ &  & 3g/4g/wifi & Disabled remote tracking system after theft & SD & 6,3,3,6,4 (4.4) & RW \\ 
\cline{2-9}  
transmission) &  &  & $\bullet$ & 3g/4g/wifi & Override of failsafe protections to reactivate vehicle & STE & 5,5,5,7,6 (5.6) & R \\ 
\hline
\hline
EPS (Steering) & $\bullet$ &  &  & Any node & EPS deactivation through compromised CAN node. & STD & 5,5,5,6,7 (5.6) & R \\ 
\hline
\hline 
Engine & $\bullet$ &  &  & Sensors & Deactivation through compromised sensor & STD & 6,5,4,7,5 (5.4) & R \\ 
\hline
\hline 
 & $\bullet$ &  &  & EV-ECU, Sensors & Critical component modification during operation & STIDE & 7,5,5,9,4 (6.0) & R \\
\cline{2-9} 
3G/4G/WiFi &  & $\bullet$ &  & Infotainment system & Privacy attack using modified radio firmware & TIE & 7,5,5,6,5 (5.6) & R \\ 
\cline{2-9} 
 &  &  & $\bullet$ & Emergency, door locks & Prevent operation of failsafe comms by & TDE & 6,6,7,8,6 (6.6) & RW \\ 
\cline{2-5} \cline{7-9}
 &  &  & $\bullet$ & Sensors, Airbags & disabling modem. & TDE & 6,6,7,8,6 (6.6) & R \\ 
\hline
\hline 
Infotainment & $\bullet$ &  &  & Media player browser & Exploit to gain access to higher control level & STE & 7,5,6,8,6 (6.4) & R \\ 
\cline{2-9}  
System &  &  & $\bullet$ & Sensors, EV-ECU & Modification of vehicle status, GPS, speed, etc & STR & 3,5,6,4,5 (4.6) & R \\ 
\hline
\hline 
Door locks & $\bullet$ &  &  & 3g/4g/wifi, Manual open & Unlock attempt while in motion & TDE & 8,5,3,8,5 (5.8) & R \\ 
\cline{2-9} 
 &  &  & $\bullet$ & 3g/4g/wifi & Lock mechanism triggered during accident & TDE & 8,6,7,8,5 (6.8) & W \\ 
\hline
\hline 
Safety  & $\bullet$ &  &  & Sensors & False triggering of failsafe mode to unlock vehicle & STE & 7,4,5,8,4 (5.6) & R \\ 
\cline{2-9} 
Critical &  &  & $\bullet$ & Sensors & Disable alarm and locking system to allow theft & TE & 9,4,5,9,4 (6.2) & W \\ 
\hline 
\end{tabular}
\begin{tablenotes}
\item \ding{118} = Normal ; \ding{74} = Diagnostics ; \ding{89} = Fail-safe

\item \textbf{STRIDE}: S=\textit{Spoofing} ; T=\textit{Tampering} ; R=\textit{Repudiation} ; I=\textit{Information Disclosure} ; D=\textit{Denial of service} ; E=\textit{Elevation of Privilege}.

\item \textbf{DREAD}: D=\textit{Damage} ; R=\textit{Reproducibility} ; E=\textit{Exploitability} ; A=\textit{Affected Users} ; D=\textit{Discoverability}.

\end{tablenotes}
\label{table:car-threats}
\end{table*}

For following threat modelling processes within the vehicle, the critical assets we have chosen to demonstrate this example are the \textit{Electronic vehicle} (EV)-ECU, \textit{electronic power steering} (EPS), Engine, 3G/4G/WiFi communications, infotainment, door locks and safety critical devices that fall under ASIL safety levels A-D, such as the immobiliser, air-bags and brake control. For the specified vehicle modes, potential threats and corresponding entry points have been identified in Table~\ref{table:car-threats}.
STRIDE threat analysis has been conducted to understand how exploited threats can affect the system, while DREAD risk-assessment has been used to analyse and quantify the likelihood of occurrence and severity of the damage in event of successful attack. 
In this case study, policies have been defined with \textit{read} or \textit{write} permissions. However, complex behavioural or situation aware policies may also be derived. Policies can be dynamically updated by the \textit{Platform Security Manager}.

\subsection{Threat and Security Modelling}

Figure~\ref{fig:can-controller} shows a block diagram of several CAN nodes connected through the CAN bus. It also shows the possible location of insider attacks caused by a compromised node or malicious modules introduced to the CAN bus. In this regard, a series of assets and corresponding threats are identified in Table~\ref{table:car-threats}. 

\subsubsection{Guideline-based security model}
Within a standard security modelling approach, the threat scenarios in Table~\ref{table:car-threats} can be approached using a guideline-based security model.
As an example within the EV-ECU and infotainment threat scenarios, the following security model guidelines may be adhered to during development.

\begin{itemize}
	\item EV-ECU: Limitation of physical access to CAN-bus and allow only authorised devices to connect. Allow read-only access on diagnostic port.
	\item Infotainment system: Provide immediate system updates when vulnerabilities are discovered.
	\item Infotainment system: Employ software protections to prevent unauthorised software installation
\end{itemize}

As a consequence of adopting a guideline-based approach, the OEM would have to redesign relevant applications in line with new requirements before redeployment. In severe cases, a recall may be required to rectify issues. Alternatively, the OEM may patch it through an update and integrate fixes in the next product cycle.

\subsubsection{Policy-based security model}
Based on the proposed security modelling, new policies can be introduced during the life-cycle of the embedded device, in case of a threat being discovered. This is performed through a \textit{trusted software} update of the \textit{platform security manager}.
This process is significantly faster and less costly to implement than a product redesign or recall. 
Table~\ref{table:car-threats} describes read and write policies that can be enforced at peripheral entry points. Fine-grained policies can also include:

\begin{itemize}
	\item Infotainment system: Prevent software installations initiated from the media display.
	\item Infotainment system: Enforce system command access using SELinux-based policies.
	\item CAN nodes: Enforce CAN ID verification at read and write filters within controller (Section~\ref{subsec:policy-enforcement}).
\end{itemize}

By utilising the policy-based approach, the OEM is not limited to addressing threats during design only. Upon detection of a new threat, policies can be defined to mitigate the attack
and be distributed through a policy update. The modelling, implementation, testing, verification and deployment process has potential to be more effective than the necessary actions within a standard guideline-based approach.

%% file: policy-enforcement.tex
\subsection{Pro-active Policing and Policy enforcement}
\label{subsec:policy-enforcement}

The concept of enforcement is a vital aspect of the proposed policy-based security modelling approach. The CAN-bus presents a security challenge, being vulnerable to
attack as shown in Figure~\ref{fig:can-controller}.
\begin{figure}[t]
\centerline{\includegraphics[scale=0.51]{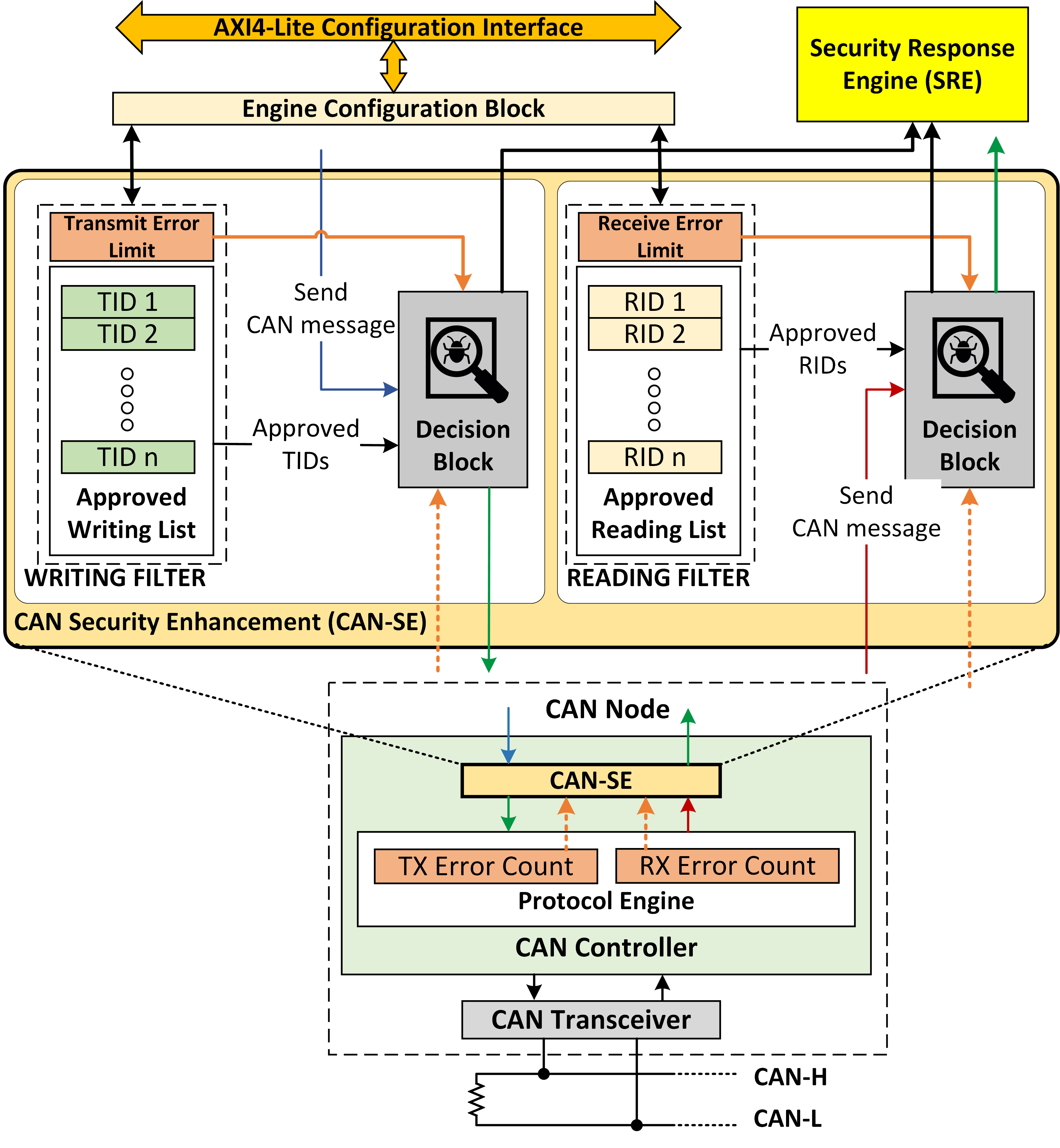}}
\caption{Block diagram of CAN node with integrated hardware-based policy engine that filters incoming and outgoing CAN messages.}
\label{fig:HPE}
\end{figure}
Within a regular CAN node, the \textit{CAN controller} utilises a software-based filter, programmed using an application and as such may be vulnerable to software layer attacks, such as firmware modification or signal-level error inducing attacks, that cause eventual shut-down of CAN-based devices~\cite{Lyu2016}.
A \textit{CAN Security Enhancement} (CAN-SE) is shown in Figure~\ref{fig:HPE}. It monitors issued read/write CAN messages, filtering them based on message IDs using hardware-based \textit{read/write filters} to detect malfunction or malicious activity. 
It consists of the following hardware components:

\begin{itemize}


\item \textit{Approved read/write lists:} Hardware components hold a list of approved CAN messages IDs to block malicious CAN messages originating either from a compromised or an introduced malicious node.

\item \textit{Error Limit Check:} Hardware-based transmit/receive components measure the number of errors introduced within the CAN bus network, notifying the presence of a malicious node
to the SRE.

\item \textit{Decision Block:} This references an approved list of message IDs, compares it to \textit{issued/received messages} and grants or blocks access as shows in Figure.~\ref{fig:HPE}.
\end{itemize}

Once the sanity check is completed, CAN messages can be either be used by the local CAN node (read) or sent (write) to the CAN bus to be utilised by other nodes thus, suppressing insider and outsider attacks to the CAN bus.

%% file: Conclusion.tex
\section{Conclusion and future work}
\label{sec:conclusion}

The paper has discussed technological advancements in embedded system architectures and the opportunities they bring to vehicle manufacturers to consolidate diverse intelligent vehicular services into smaller, flexible and integrable vehicular electronics and domain controllers. However, digital-physical convergence, mixture of critical and non-critical, secure and non-secure vehicle services and shortcomings in existing embedded security approaches present significant challenges. Serious security and safety concerns are inevitable particularly where malicious events or malfunction of intelligent vehicle services occur. This work has proposed a novel pro-active policing and policy-based security platform architecture for next-generation in-vehicle domain controllers, providing a pro-active fail-over mechanism for intelligent vehicle services. This security platform continuously monitors system bus-level communications to detect unexpected and malicious behaviours and pro-actively respond against such activities to control, confine and minimise physical damage and harm to passengers and pedestrians. The proposed architecture has been prototyped and verified on Avnet UltraZed and Zedboard development boards.

An agile policy-based security modelling approach has been proposed to model and integrate the proposed policy-based security approach using existing security modelling methods. A realistic \textit{connected vehicle} case study has been presented to highlight and demonstrate its flexibility and how it can be integrated within the existing secure development life-cycle and security design practices. A detailed domain-specific threat modelling of a connected vehicle case study has been conducted using STRIDE and DREAD risk assessment models to derive security policies and enforced by the platform. In contrast to a guideline-based approach, this provide opportunities to realise adaptable system security, implementing diverse use case security requirements without the need to redesign the underlying security architecture.

The presented work will be extended to gain and achieve contextual-based vehicle access control, thus enabling enforcement of system-level behavioural policies.